\documentclass[twocolumn,showpacs,superscriptaddress,preprintnumbers,amsmath,amssymb,prc]{revtex4-1}

\usepackage{graphicx}
\usepackage{dcolumn}
\usepackage{bm}
\usepackage{hyperref}
\usepackage{soul} 
\usepackage{color, xcolor} 
\soulregister\cite7
\soulregister\ref7

\begin{document}

\title{New insight into the $N/Z$ and mass equilibration in heavy-ion collisions}

\author{Yu Yang}
\affiliation{Sino-French Institute of Nuclear Engineering and Technology, Sun Yat-sen University, Zhuhai 519082, China}

\author{Zehong Liao}
\affiliation{Sino-French Institute of Nuclear Engineering and Technology, Sun Yat-sen University, Zhuhai 519082, China}

\author{Zepeng Gao}
\affiliation{Sino-French Institute of Nuclear Engineering and Technology, Sun Yat-sen University, Zhuhai 519082, China}

\author{Long Zhu}
\email[Corresponding author, ]{zhulong@mail.sysu.edu.cn}
\affiliation{Sino-French Institute of Nuclear Engineering and Technology, Sun Yat-sen University, Zhuhai 519082, China}

\affiliation{Guangxi Key Laboratory of Nuclear Physics and Nuclear Technology, Guangxi Normal University, Guilin 541004, China}

\author{Jun Su}
\affiliation{Sino-French Institute of Nuclear Engineering and Technology, Sun Yat-sen University, Zhuhai 519082, China}
\affiliation{Guangxi Key Laboratory of Nuclear Physics and Nuclear Technology, Guangxi Normal University, Guilin 541004, China}

\author{Cheng Li}
\affiliation{Guangxi Key Laboratory of Nuclear Physics and Nuclear Technology, Guangxi Normal University, Guilin 541004, China}
\affiliation{Department of Physics, Guangxi Normal University, Guilin 541004, China}

\date{\today}

\begin{abstract}
The dynamics of $N/Z$ and mass equilibration are investigated in the reactions $^{112,124}$Sn + $^{239}$Pu by employing the isospin-dependent quantum molecular dynamics model. It is found that $N/Z$ and mass equilibration take place at different collision stages. The $N/Z$ relaxation is observed in the approaching phase (from first contact to deepest contact) with a very short time, whereas interestingly we find for the first time that mass equilibration only takes place in the separation phase (from the deepest contact to re-separation), which are explained by investigating the dynamical asymmetry between the approaching and separation phases. The mass equilibration also could be clarified with a dynamical potential energy surface. Our results provide a new insight into the equilibration dynamics of the quantum systems.

\end{abstract}
 
\maketitle

The equilibration mechanism of quantum systems has been an open problem for more than a decade \cite{eisert2015quantum,gring2012relaxation,simenel2020timescales}, which is connected to the temporal evolution of quantum systems in heavy-ion collisions. 
The dynamical properties of equilibration in heavy-ion collisions are known to be strongly affected by the transfer of nucleons leading to the exchange of mass and charge as well as the dissipation of energy \cite{schroderDampedHeavyIonCollisions1977, huizenga1976energy,williams2018exploring,jiang2013dynamics,feldmeier1984particle,schroder1978mechanisms,wilcke1980bombardingenergy,wollersheim1981bi}. The neutron-to-proton ratio ($N/Z$) equilibration has attracted a lot of attention due to the intriguing experimental observation that $N/Z$ equilibration is a fast equilibration mode in comparison to mass asymmetry relaxation \cite{liao2023dynamics,jedele2017characterizing,krolas2010dynamical,kratz1977chargeasymmetry,hernandez1981quantal,mathewsNe20Induced1982,freiesleben1984nzequilibration}. This raises a question of what are the differences between the mechanisms of $N/Z$ and mass equilibration.

Several studies suggest that the quantal behaviors, such as the giant dipole resonance in deep inelastic collisions (DIC) account for the very quick relaxation of $N/Z$ to equilibration \cite{moretto1979giant,brosa1978time,berlanger1979studya,hofmann1979theoretical,wuDynamicalDipoleRadiation2010,iwata2010suppression}. This is because $N/Z$ equilibration is much faster than the one connected to the energy dissipation. While it is commonly assumed that the statistical mechanism is responsible for the mass equilibration \cite{schroderDampedHeavyIonCollisions1977}. Usually, the equilibration and dissipation processes in DIC are investigated based on the properties of the products after collisions overlooking the dynamical effects during the collision process. Therefore, the mechanisms of the temporal evolution of $N/Z$ and mass asymmetry are still not clear, which hinders the understanding of the dynamics of the equilibration process in quantum systems.

In present Letter, we aim to clarify the dynamics of $N/Z$ and mass equilibration in DIC. More specifically, we investigate the dynamical properties of equilibration in collision stages of approaching (from first contact to deepest contact) and separation (from the deepest contact to re-separation) by comparing the reactions $^{112,124}$Sn + $^{239}$Pu within the framework of the isospin-dependent quantum molecular dynamics (IQMD) model \cite{zhang1999isospin,chen1999isospin,chen2000isospin}.

\begin{figure*}[htp]
    \centering
    \includegraphics[width=.85\linewidth]{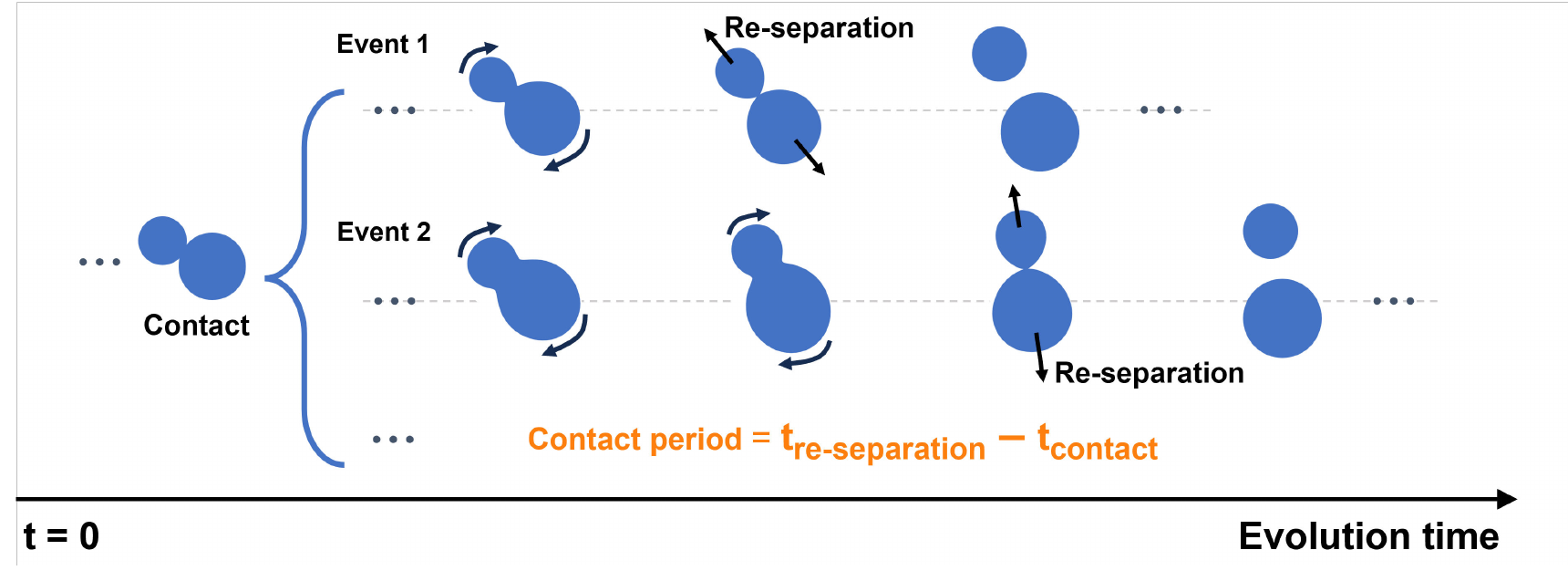}
    \caption{The schematic of the evolution process of a dinuclear system. The contact period is defined as the duration from first contact to re-separation.
    }
    \label{fig:Schematic diagram}
\end{figure*}

The IQMD model is a microscopic transport model, in which each nucleon is described by a coherent state of a Gaussian wave packet singly peaked at the center of the particle \cite{aichelin1991quantum}.
In the IQMD model, nucleons move in the self-consistent mean ﬁeld generated by all other nucleons and the propagation satisﬁes the Hamiltonian canonical equations of motion: 
\begin{eqnarray}
    \dot{\textbf{r}}_i= \frac{\partial H}{\partial \textbf{p}_i} 
    ,~~\dot{\textbf{p}}_i= -\frac{\partial H}{\partial \textbf{r}_i}\label{eq_motion}.
\end{eqnarray}
The Hamiltonian of the system is written as the sum of the kinetic energy $T=\sum_i \frac{p_i^2}{2m}$ and effective interaction potential energy:
\begin{eqnarray}
    H = T + U_{\mathrm{coul}} + U_{\mathrm{loc}}\label{eq_hamilton},
\end{eqnarray}
where $m$ represents the nucleon mass, and $U_{\mathrm{coul}}$ is the Coulomb interaction potential energy, which is written as the sum of the direct and the exchange contribution, with the latter being taken into account in the Slater approximation \cite{slater1951simplification}
\begin{eqnarray}
    U_{\mathrm{coul}} = \frac{e^2}{2}\iint \frac{\rho_{\mathrm{p}}(\textbf{r})\rho_{\mathrm{p}}(\textbf{r}^{\prime})}{\lvert \textbf{r}-\textbf{r}^{\prime} \rvert} d\textbf{r}d\textbf{r}^{\prime} 
    \notag
    \\
    - \frac{3e^2}{4}\left(\frac{3}{\pi}\right)^{1/3}\int \rho_{\mathrm{p}}^{4/3}d\textbf{r}\label{eq_coulomb},
\end{eqnarray}
where $\rho_{\mathrm{p}}$ is the proton density of the system. $U_{\mathrm{loc}}$ is the nuclear interaction potential energy, which is obtained from the integration of the Skyrme energy density functional $U_{\mathrm{loc}} = \int{V_{\mathrm{loc}}(\textbf{r})d\textbf{r}}$ \cite{wang2002improved}. The nuclear interaction potential density $V_{\mathrm{loc}}$ can be written as 
\begin{eqnarray}
    V_{\mathrm{loc}} = \frac{\alpha}{2}\frac{\rho^2}{\rho_0} 
    + \frac{\beta}{\gamma+1}\frac{\rho^{\gamma+1}}{\rho_0^{\gamma}} 
    + \frac{g_{\mathrm{sur}}}{2\rho_0}(\nabla\rho)^2 
    \notag
    \\
    + \frac{C_{\mathrm{s}}}{2\rho_0}[\rho^2-\kappa_{\mathrm{s}}(\nabla\rho)^2]\delta^2 
    + g_{\mathrm{\tau}}\frac{\rho^{\eta+1}}{\rho_0^{\eta}}\label{eq_skyrme},
\end{eqnarray}
where $\rho = \rho_n + \rho_p$ is the nucleon density, and $\delta = (\rho_{\mathrm{n}} - \rho_{\mathrm{p}})/(\rho_{\mathrm{n}} + \rho_{\mathrm{p}})$ is the isospin asymmetry. The density distribution in the coordinate space $\rho(\mathbf{r})$ is given by 
\begin{align}
    \rho(\mathbf{r}) = \sum_i \frac{1}{(2\pi\sigma_{\mathrm{r}}^2)^{3/2}}
    \exp\left[-\frac{(\mathbf{r} - \mathbf{r}_i)^2}{2\sigma_{\mathrm{r}}^2}\right] \label{eq_rho}
\end{align}
where $\sigma_{\mathrm{r}} =0.88 + 0.09A^{1/3}$ fm is  the Gaussian wave-packet width. The parameters in Eq. (\ref{eq_skyrme}) are related to the standard Skyrme interaction parameters, with $\alpha = -356$ MeV, $\beta = 303$ MeV, $\gamma = 7/6$, $g_{\mathrm{sur}} = 7$ MeV fm$^2$, $g_{\mathrm{\tau}} = 12.5$ MeV, $\eta = 2/3$, $C_{\mathrm{s}} = 32$ MeV, $\kappa_{\mathrm{s}} = 0.08$ fm$^2$, and $\rho_0 = 0.165$ fm$^{-3}$. 

To describe the fermionic nature of the $N$-body system and to improve the stability of an individual nucleus, the phase space occupation constraint method is adopted \cite{papa2005constrained}. If the phase space occupation number is greater than 1 for $i$ th nucleon, the momentum of $i$ th nucleon is randomly changed by a series of two-body elastic scatterings between $i$ th nucleon and its surrounding partners, due to the random change in momentum being in the center-of-mass frame of the two nucleons, the momentum and kinetic energy of the system are conserved. The initial center-of-mass distance between the projectile and target is 30 fm. The impact parameter $b$ is corrected considering a Coulomb trajectory, which depends on the charge and mass of the projectile-target combination as well as the incident energy \cite{maruyama1990quantum}.

Once the system overcomes the Coulomb barrier, density overlap arises due to attractive nuclear forces,  initiating the formation of a neck. It is known that the dynamical properties of the dinuclear complex are strongly affected by the multiple nucleon transfer occurring between the two parts of the system. The formation of the neck results in a significant exchange of mass, charge, kinetic energy, linear, and angular momentum between the two nuclei \cite{zhao2009mass,jiang2013dynamics, li2016multinucleon,li2020production,li2018production}. To divide the projectile-like fragment (PLF) and target-like fragment (TLF) for the dinuclear system with a neck, we introduce a window plane along the $z$-axis of the symmetry axis in the rotating frame, following the segmentation method as in \cite{yilmaz2011nucleon,yilmaz2014nucleon,feng2023contributions,feng2024microscopic}. We can easily obtain the center-of-mass position $R_i$ and momentum $P_i$ of the fragments:
\setcounter{equation}{6}
\begin{align}
    &\text{$R$}_i
    = \frac{1}{A_i} \int_{V_i} d\textbf{r}~z
    \int d\textbf{p}~
    f(\textbf{r},\textbf{p})
    \tag{6a}\label{eq_R},
    \\
    &\text{$P$}_i 
    = \int_{V_i} d\textbf{r} \int d\textbf{p}~p_{z}
    f(\textbf{r},\textbf{p})
    \tag{6b}\label{eq_P},
\end{align}
where $f(\mathbf{r},\mathbf{p})$ represents the one-body distribution function in phase space \cite{wang2002improved}, and $V_i$ represents the subspace with $i$ referring to PLF and TLF, respectively. $A_i = \int_{V_i} d\mathbf{r} \int d\mathbf{p} f(\mathbf{r},\mathbf{p})$ is the mass of each subsystem and $p_z$ is the momentum component along the symmetry axis.
Then, the mass parameter of each part is obtained as $m_i = P_i/\dot R_i$. Thus, the relative distance $R_{\mathrm{c.m.}}(t)$ and associated momentum $P(t)$ are given by :
\setcounter{equation}{7}
\begin{align}
     &R_{\mathrm{c.m.}}(t) = R_{\mathrm{TLF}} - R_{\mathrm{PLF}},\tag{7a}\label{eq_Rcm}\\
     &P(t) = \frac{m_{\mathrm{TLF}}P_{\mathrm{PLF}} - m_{\mathrm{PLF}}P_{\mathrm{TLF}}}{m_{\mathrm{PLF}}+m_{\mathrm{TLF}}}.\tag{7b}\label{eq_P_realtive}
\end{align}

\begin{figure}[htp]
    \centering
    \includegraphics[width=8.0cm]{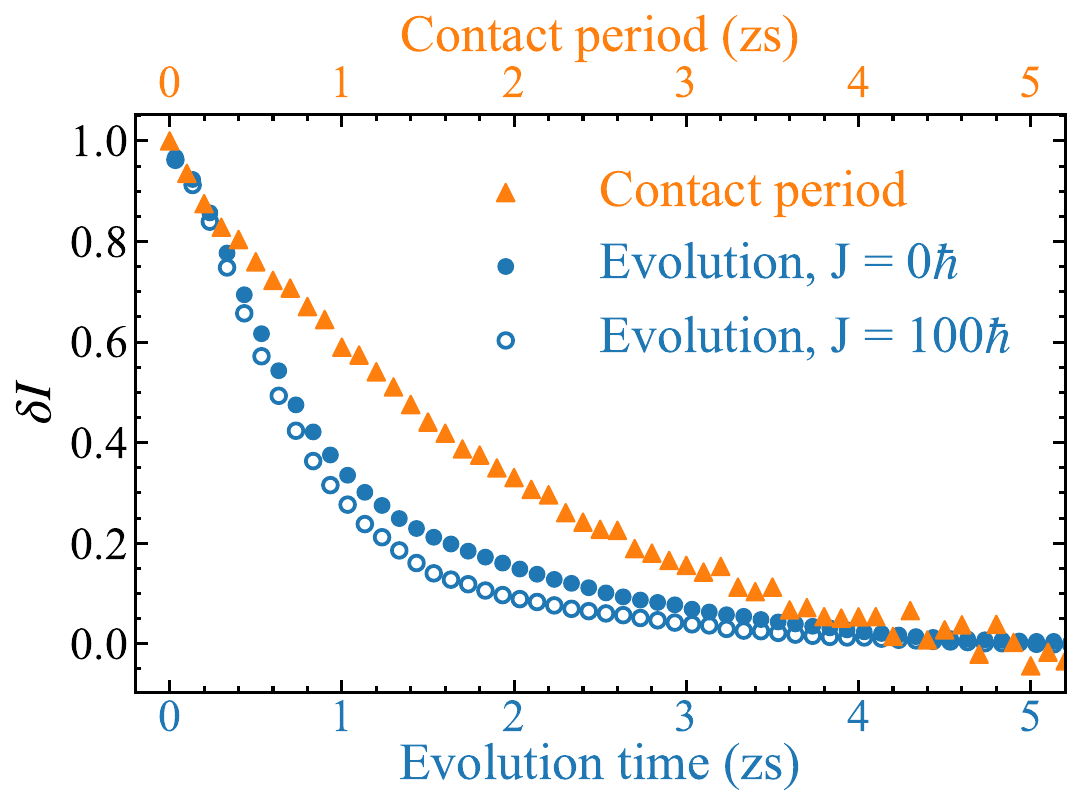}
    \caption{The $\delta I$ values for the primary PLF as a function of the contact period (orange) and the evolution time (blue). For the latter, the solid (open) dots denote the incident angular momentum $J=$ 0$\hbar$ (100$\hbar$), and the evolution time is set to 0 as the initial contact moment of the system.
    }
    \label{fig:NZ equilibration}
\end{figure}

The timescale of $N/Z$  equilibration is usually investigated and estimated by examining the dependence of the $N/Z$ ratio of the products over the contact period \cite{kratz1977chargeasymmetry,jedele2017characterizing,umar2017transport,rehm1979timea}.
In Fig. \ref{fig:Schematic diagram}, we present a diagram of the evolution process during the collision to illustrate the difference between the ``contact period'' and ``evolution time''. The contact period is the duration from the first contact of the colliding partners to their re-separation, which depends on factors such as the impact parameters, the colliding system, and the incident energy. Therefore, analyzing the $N/Z$ equilibration based solely on the contact period overlooks the dynamical effects present throughout the entire collision process , especially the different dynamical properties between approaching (from first contact to deepest contact) and separation (from the deepest contact to re-separation) processes. 

To elucidate the dynamical effects of the $N/Z$ equilibration, the time evolution of $\delta I = [I(t) - I_{\infty}]/(I_0 - I_{\infty})$ values in the reaction $^{112}$Sn + $^{239}$Pu at relative angular momentum $J=$ 0 and 100$\hbar$ are depicted in Fig. \ref{fig:NZ equilibration}. $I_0$ and $I_{\infty}$ denote the initial and expected saturation values of $N/Z$ for the projectile, respectively. The values of $N/Z$ for $^{112}$Sn and $^{239}$Pu are 1.24 and 1.55, respectively. From the fitting equation $\delta I = y_0 + A_0 \exp(-t / \tau)$, the characteristic time of equilibration (CTE) $\tau$ could be extracted as 0.9 and 0.7 zs for $J=$ 0 and 100$\hbar$, respectively. Additionally, we show the variation of $\delta I$ with contact period which strongly depends on the impact parameters \cite{umar2017transport}. However, interestingly, the CTE obtained from the contact period is 2.2 zs, which is much longer than that from the evolution time. One would like to ask which way for evaluating the CTE of $N/Z$ is reasonable.

\begin{figure}[htp]
    \centering
    \includegraphics[width=8.5cm]{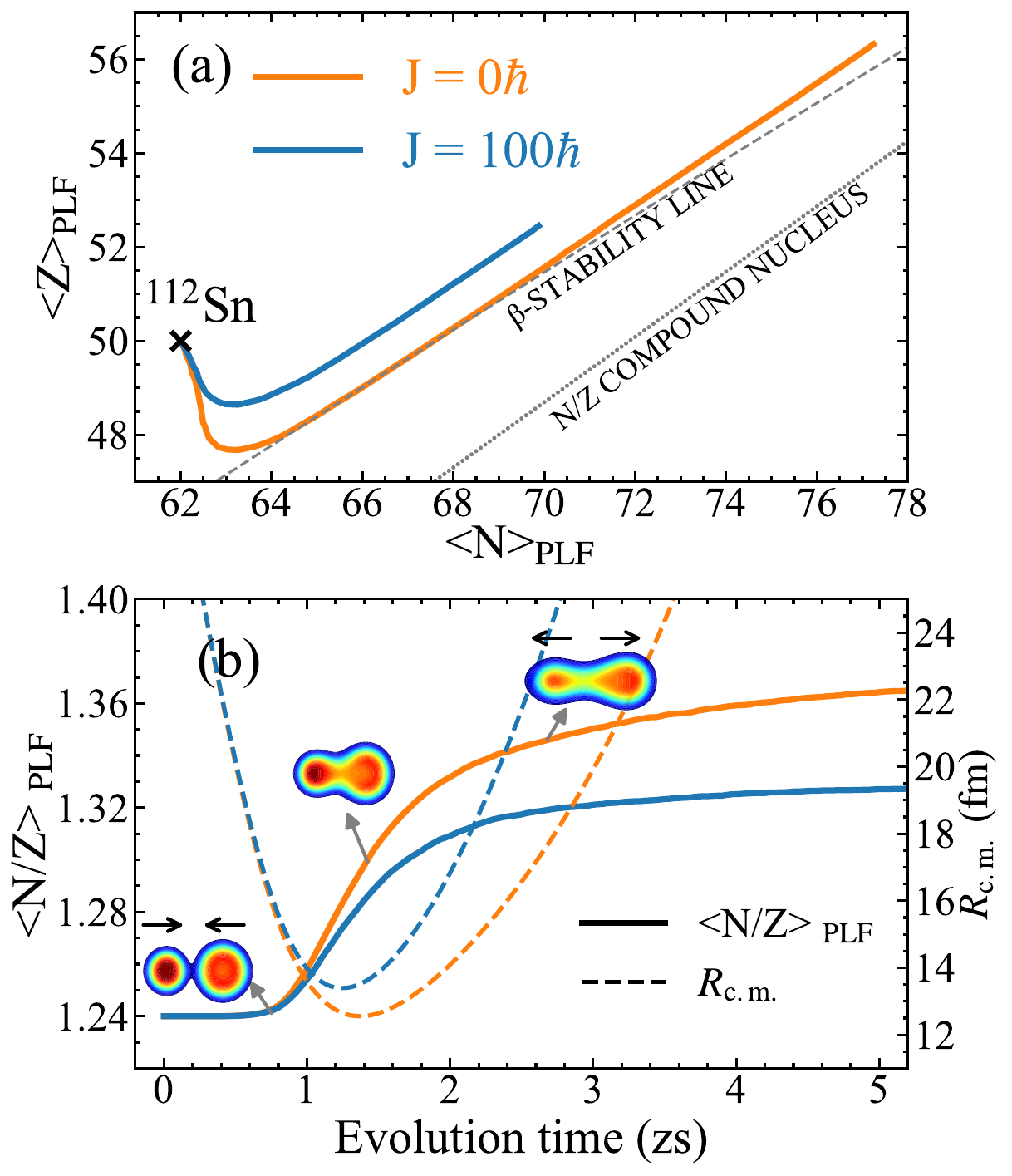}
    \caption{(a) Mean drift paths of the PLF in the ($N$,$Z$) plane for the $^{112}$Sn + $^{239}$Pu system with initial angular momenta $J=$ 0 and 100$\hbar$. The initial position of $^{112}$Sn is marked with a cross. (b) The $N/Z$ of the PLF (thick) and center-of-mass distance $R_{\mathrm{c.m.}}$ (dashed) as a function
    of the evolution time. Density contour plots illustrate key stages such as just in contact, deepest in contact, and in the process of separation, the black arrows represent the direction of movement.}
    \label{fig:mean path}
\end{figure}

Fig. \ref{fig:mean path}(a) illustrates the dynamical trajectories in the ($N$, $Z$) plane for the initial angular momenta $J=$ 0 and 100$\hbar$. The incomplete $N/Z$ equilibration is noticed. 
It is also evident that the $N/Z$ evolution stems from the disparate charge asymmetries of the initial reaction partners and exhibits similar behavior across different angular momenta. 
However, notably, the paths do not overlap, which means that although the $N/Z$  evolves rapidly, the saturation of $N/Z$ relaxation is not simply dependent on the contact time. In Fig. \ref{fig:mean path}(b), we show the variation of $N/Z$ and  $R_{\mathrm{c.m.}}$ with evolution time. One can see that the $N/Z$ evolution of the system approaches saturation as it reaches its deepest contact. In other words, the $N/Z$ equilibration takes place in the approaching stage before the deepest contact. In fact, although it has been noted that the $N/Z$ evolution is fast by following the tendency of the gradient of the potential energy surface (PES) to minimize the potential energy, it is also implied that the system preferentially follows the same path to reach $N/Z$ equilibration, even in different cases. However, our results show that it is likely controlled by factors beyond the relevant static potentials, most likely by dynamical effects. As pointed out in \cite{schull1981influence}, it may also be related to the contact area between the two nuclei.

Furthermore, to better investigate the transfer of nucleons during the approaching and separation phases of collisions, the concept of drift coefficients has been introduced \cite{yilmaz2014nucleon}:
\begin{align}
    v_{\mathrm{k}}(t) 
    &= \frac{dA_{\mathrm{k}}}{dt} \notag \\
    &= \int d{x}d{y} \int d\mathbf{p} 
      \frac{p_z-p_{z0}}{m}
      f_{\mathrm{k}}(\mathbf{r},\mathbf{p})\mid_{z = z_s}.\label{eq_flux}
\end{align}
Here $m$ signifies the nucleon mass, and $p_z$ represents the momentum component along the symmetry axis, while $p_{z0}$ represents the momentum of the window motion. The nucleon ﬂux is evaluated over the window plane deﬁned by $z$ = $z_s$. The subscript k denotes the four components: the proton and neutron of the projectile and the target. In addition, the positive direction of the momentum is from the projectile to the target. The drift coefficients provide microscopic information about the collision processes, including approaching and separation. Due to the small amplitude oscillations of the window positions, the drift coefficients show small temporal variations.

\begin{figure}[htp]
    \centering
    \includegraphics[width=8.5cm]{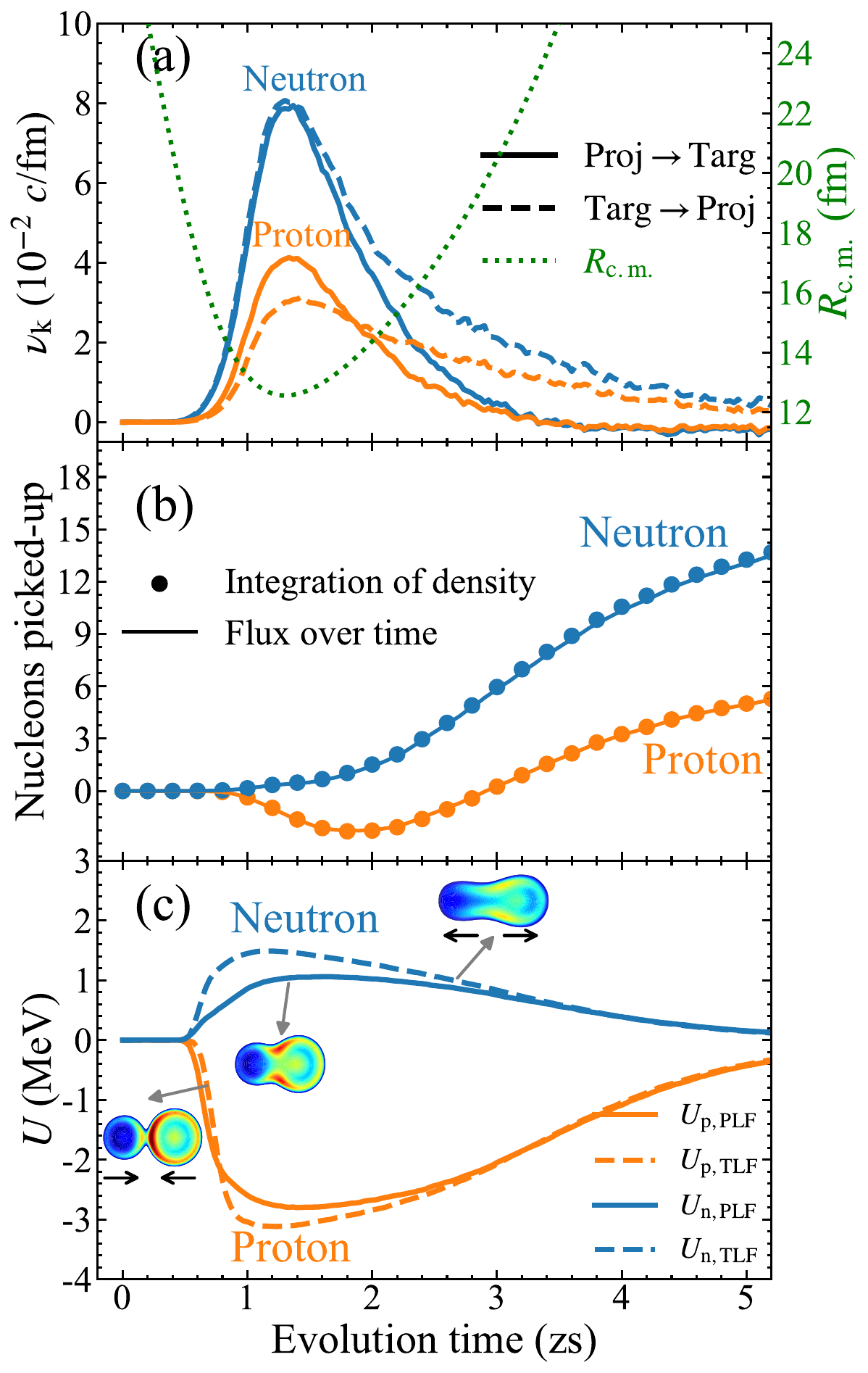}
    \caption{(a) The proton drift $\nu_{\mathrm{p}}$ (orange line), neutron drift $\nu_{\mathrm{n}}$ (blue line), and center-of-mass distance $R_{\mathrm{c.m.}}$ (dotted line) for the $^{112}$Sn + $^{239}$Pu as functions of the evolution time at initial angular momentum of $J=$ 0$\hbar$. (b) The particle number of nucleon transfer from the target to projectile as functions of the evolution time. The solid dots are calculated by direct density integration, while the solid lines are obtained by integrating the nucleon flux over time. (c) The average energy for neutrons and protons at the neck region as functions of the evolution time, accompanied by contour plots depicting isospin asymmetry $\delta = (\rho_{\mathrm{n}}-\rho_{\mathrm{p}})/\rho$ as in Fig. \ref{fig:mean path} (b).}
    \label{fig:drift}
\end{figure}

From Fig. \ref{fig:drift}(a), it can be seen that the drift coefficients of the nucleons increase dramatically during the approaching phase of the two colliding nuclei, peaking around their deepest contact at approximately 1.3 zs. Subsequently, the drift coefficients exhibit a gradual decrease, resulting in an overall non-Gaussian shape. Additionally, the drift coefficients exhibit a rather long tail over very large distances. This is attributed to the formation of a large neck and the presence of strong dynamical fluctuations in head-on collisions \cite{zhao2009mass,wang2016new}.
Notably, the proton drift coefficients initially experience a much slower increase due to the presence of Coulomb barriers, hindering proton transfer in the early stages of the collision \cite{li2019isospin}. Surprisingly, in the approaching phase, despite fast increase of the neutron drift coefficients, the resulting amount of the net transfer is small due to the very close values of $v$ for both directions. However, the proton drift coefficient in projectile-to-target direction increases more significantly than that in the opposite one, which results in the net transfer of protons from the projectile to the target, and then causes the intense relaxation of $N/Z$ toward equilibration. Around 1.8 zs, one can see that the proton drift coefficient curves for two directions intersect with each other. This intersection indicates a shift in net transfer direction, as shown in \ref{fig:drift}(b). On the other hand, the neutrons tend to transfer from target to projectile, especially after the deepest contact, and the drift coefficients gradually decrease during separation. These observations further support that the $N/Z$ equilibration predominantly occurs during the approaching phase of the colliding nuclei, while during the separation phase, the $N/Z$ ratio has already converged to its saturation value, coinciding with the occurrence of mass equilibration in the system. Furthermore, the differences in characteristic times for calculating the $N/Z$ equilibration in Fig. \ref{fig:NZ equilibration} can be attributed to the these observed dynamical asymmetry in approaching and separation processes. Therefore, the CTE of $N/Z$ should be evaluated from the view point of evolution time rather than contact period.

Why the $N/Z$ evolution of the system almost saturates near the point of the deepest contact? As mentioned earlier in this Letter, the neck plays an important role in DIC, and the isospin transport of the system is closely related to the symmetry and Coulomb energies. This implies that the characteristics of the neck region may dictate the flow of nucleons during most of the dynamical evolution. Fig. \ref{fig:drift}(c) illustrates the time evolution of the average potential energies for neutrons ($U_{\mathrm{sym,n}}$) and protons ($U_{\mathrm{sym,p}} + U_{\mathrm{Coul,p}}$) in the neck region. Here, $U_{\mathrm{n,p}}$ can be obtained directly from the integral of the potential energy density functional given in Eq. (\ref{eq_coulomb}) and Eq. (\ref{eq_skyrme}). The neck region is deﬁned as a cylindrical shape along the symmetry orientation extending up to 3 fm. As can be seen from the Fig. \ref{fig:drift}(c), initially, $\Delta U_{\mathrm{p}}=U_{\mathrm{p,PLF}}-U_{\mathrm{p,TLF}}$ is positive and $\Delta U_{\mathrm{n}}=U_{\mathrm{n,PLF}}-U_{\mathrm{n,TLF}}$ is negative, indicating that the PLF has a tendency to gain neutrons and lose protons. 

\begin{figure*}[htp]
    \centering
    \includegraphics[width=.85\linewidth]{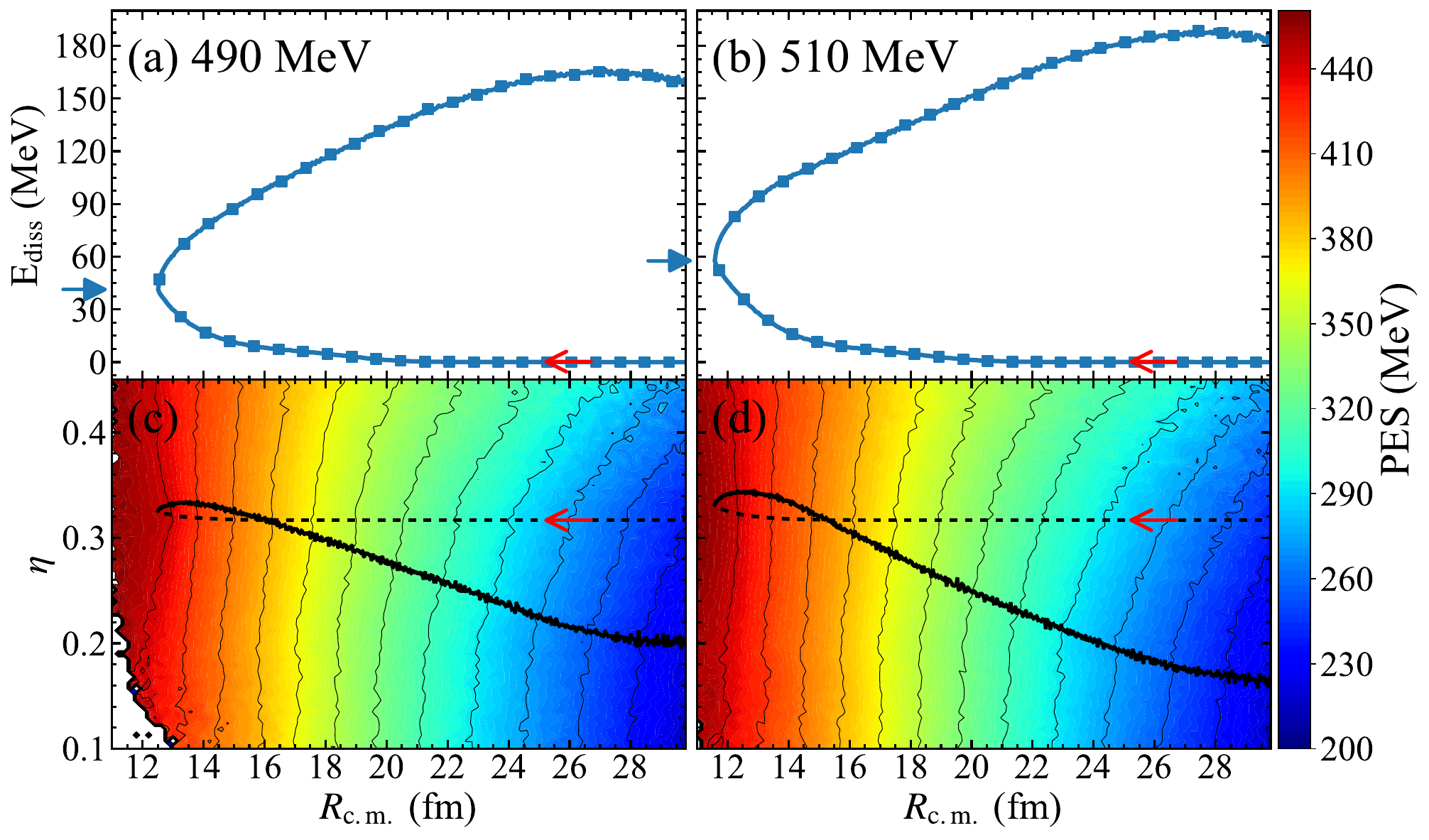}
    \caption{The dissipated energy (top) and mass asymmetry (bottom) for $^{124}$Sn + $^{239}$Pu at two different energies as a functions of center-of-mass distance $R_{\mathrm{c.m.}}$ at initial angular momentum of $J =$ 0$\hbar$. The corresponding dissipated energy at the deepest contact is marked with a blue arrow. In the bottom panel, the dashed line indicates the approaching process, while the thick line indicates the separation process. The contours show the DyPES in the separation stage.
    }
    \label{fig:mass_Ediss}
\end{figure*}

In Fig. \ref{fig:drift}(a), from the asymmetric nucleon drift properties before and after the deepest contact point, one can observe that the mass relaxation mainly takes place at the separation stage. Due to this delayed relaxation, one might wonder if there is a threshold excitation energy for the mass equilibration. Using the IQMD model, we can extract the dissipated energy based on energy conservation as follows:
\begin{align}
    &E_{\mathrm{diss}}(R_{\mathrm{c.m.}}) = E_{\mathrm{c.m.}} - \frac{P^2}{2\mu(R_{\mathrm{c.m.}})} - V(R_{\mathrm{c.m.}}),\label{eq_Ediss}\\
    &V(R_{\mathrm{c.m.}}) = E_{\mathrm{tot}}(R_{\mathrm{c.m.}}) - E_{\mathrm{PLF}}(R_{\mathrm{c.m.}}) - E_{\mathrm{TLF}}(R_{\mathrm{c.m.}})\label{eq_Vint}.
\end{align}
Here, the second term in Eq. (\ref{eq_Ediss}) represents the kinetic energy of the relative motion, where $P$ is obtained from Eq. (\ref{eq_P_realtive}) and $\mu(R_{\mathrm{c.m.}}) = P/\dot R_{\mathrm{c.m.}}$ is the reduced mass. $V(R_{\mathrm{c.m.}})$ is the nucleus-nucleus potential, where $E_{\mathrm{tot}}$, $E_{\mathrm{PLF}}$, and $E_{\mathrm{TLF}}$ are the total energy of the whole system, the energy of the PLF and TLF, respectively. These energies are obtained by integrating Eq. (\ref{eq_skyrme}) over the whole system, the PLF and TLF. Additionally, the kinetic-energy density functional is described by the extended Thomas-Fermi method \cite{jiang2010dynamical} in Eq. (\ref{eq_Vint}) to eliminate the contribution of collective motion.

The reaction $^{124}$Sn + $^{239}$Pu is investigated to reveal the relationship between mass asymmetry $\eta$ $[= (A_{\mathrm{TLF}}-A_{\mathrm{PLF}})/(A_{\mathrm{TLF}}+A_{\mathrm{PLF}})]$ evolution and energy dissipation. Because both nuclei have close values of $N/Z$, the analysis of the mass equilibration is not expected to be affected by the $N/Z$ equilibration. From Fig. \ref{fig:mass_Ediss}(a) and (c), it can be seen that despite the rapid dissipation of the collective kinetic energy after contact, the mass asymmetry of the system remains nearly constant, even slightly inverse mass equilibration takes place in the approaching stage. It is worth noting that, as shown in Fig. \ref{fig:drift}(a), there is also a significant exchange of nucleons, although the net number of transferred nucleons is small. During the separation phase, mass equilibration begins to develop. While the collective kinetic energy continues to dissipate at a slightly lower rate compared to the approaching phase, more kinetic energy is dissipated during the separation phase due to its longer duration. To investigate whether there exists a threshold for mass equilibration, as discussed previously, the case for higher incident energy $E_{\mathrm{c.m.}}=510$ MeV is also shown in Fig. \ref{fig:mass_Ediss}(b) and (d). At higher incident energy, although the collective kinetic energy is dissipated faster, it is intriguing to notice that the mass equilibration is still dominated by the separation phase, but with more dramatic behavior.

It has been suggested that the mass equilibration is strongly influenced by the diameter of the neck, which provides space and time for a relatively unrestricted flow of nucleons from the heavier to the lighter reaction partner. However, our calculations suggest that this may not be the sole cause of mass equilibration, since the values of $v$ for both directions are very close in the approaching phase. One possible reason for this discrepancy is that, during the approaching phase, although a large number of nucleons are exchanged and the incident kinetic energy gradually dissipates, the two nuclei remain relatively intact. The valence nucleons are mainly concentrated in the neck region shared by the two fragments. This allows $N/Z$ equilibration to occur, but may not be sufficient to cause mass drift. In contrast, during the separation phase, the incident kinetic energy is completely damped, and the nucleons in the neck are reassigned to PLF and TLF which is governed by the PES.

Explanations for the rapid $N/Z$ equilibration have primarily focused on the static PES, with limited exploration from a dynamic evolutionary perspective. In this work, in order to reveal above intriguing phenomenon, we proposed a dynamical PES (DyPES), which is a function of the relative distance $R_{\mathrm{cm}}$ and mass asymmetry $\eta$: 
\begin{align}
    U(R_{\mathrm{c.m.}},\eta) 
    &= V(R_{\mathrm{c.m.}},\eta)\\ \notag
    &+ B_{\mathrm{PLF}}(R_{\mathrm{c.m.}}) + B_{\mathrm{TLF}}(R_{\mathrm{c.m.}})
    - B_{\mathrm{pro}} - B_{\mathrm{tar}}\label{eq_PES}.
\end{align}
Here, $V(R_{\mathrm{cm}},\eta)$ represents the nucleus-nucleus potential obtained from Eq. (\ref{eq_Vint}). $B_{\mathrm{PLF}}$ and $B_{\mathrm{TLF}}$ denote the binding energies of the two fragments. Similarly, $B_{\mathrm{pro}}$ and $B_{\mathrm{tar}}$ refer to the binding energies of the projectile and the target, respectively. 
The IQMD model automatically accounts for dynamic effects and fluctuations, providing a more comprehensive representation compared to the static PES. The contours in Fig. \ref{fig:mass_Ediss}(c) and (d) display the DyPES at the separation phase for the reaction $^{124}$Sn + $^{239}$Pu at $E_{\mathrm{c.m.}}=$ 490 and 510 MeV, respectively. As observed in Fig. \ref{fig:mass_Ediss}(c) and (d), the DyPES as a whole is tilted along the lower right side, which follows the same trend as the mass asymmetry evolutionary trajectory of the system.


In summary, within the framework of the IQMD model, it is found that the CTE extracted from the view points of evolution time and contact period are quite different. To better understand the equilibrium mechanism, we investigate the dynamical features of heavy-ion collisions $^{112,124}$Sn + $^{239}$Pu near the Coulomb barrier, focusing on the temporal evolution of the $N/Z$ ratio and mass equilibration. The neutron and proton drift coefficients are investigated in the reaction $^{112}$Sn + $^{239}$Pu. Our results show that the $N/Z$ equilibration process nearly saturates at the deepest point of contact, which means the separation stage contributes little to the $N/Z$ equilibration. It is also shown that during the approaching phase, the nucleons exchange process plays a major role with very weak mass asymmetry variation. Furthermore, the mass equilibration process in $^{124}$Sn + $^{239}$Pu reaction at different energies is investigated. Intriguingly, it is found that mass equilibration only takes place in the separation phase. With the definition of the DyPES, the mass equilibration is explained that the system evolves along the fastest way to lower the potential energy. Because the $N/Z$ and mass equilibration processes do not occur throughout the whole contact period, the characteristic time of equilibration should be evaluated from the view point of time evolution rather than contact period as shown in most experimental and theoretical studies.

This work was supported by the National Natural Science Foundation of China under Grants No. 12075327; The Open Project of Guangxi Key Laboratory of Nuclear Physics and Nuclear Technology under Grant No. NLK2022-01; Fundamental Research Funds for the Central Universities, Sun Yat-sen University under Grant No. 23lgbj003; The Guangxi Natural Science Foundation Grants No. 2023GXNSFBA026008.

\bibliography{bibliography}

\begin{thebibliography}{46}%
\makeatletter
\providecommand \@ifxundefined [1]{%
 \@ifx{#1\undefined}
}%
\providecommand \@ifnum [1]{%
 \ifnum #1\expandafter \@firstoftwo
 \else \expandafter \@secondoftwo
 \fi
}%
\providecommand \@ifx [1]{%
 \ifx #1\expandafter \@firstoftwo
 \else \expandafter \@secondoftwo
 \fi
}%
\providecommand \natexlab [1]{#1}%
\providecommand \enquote  [1]{``#1''}%
\providecommand \bibnamefont  [1]{#1}%
\providecommand \bibfnamefont [1]{#1}%
\providecommand \citenamefont [1]{#1}%
\providecommand \href@noop [0]{\@secondoftwo}%
\providecommand \href [0]{\begingroup \@sanitize@url \@href}%
\providecommand \@href[1]{\@@startlink{#1}\@@href}%
\providecommand \@@href[1]{\endgroup#1\@@endlink}%
\providecommand \@sanitize@url [0]{\catcode `\\12\catcode `\$12\catcode `\&12\catcode `\#12\catcode `\^12\catcode `\_12\catcode `\%12\relax}%
\providecommand \@@startlink[1]{}%
\providecommand \@@endlink[0]{}%
\providecommand \url  [0]{\begingroup\@sanitize@url \@url }%
\providecommand \@url [1]{\endgroup\@href {#1}{\urlprefix }}%
\providecommand \urlprefix  [0]{URL }%
\providecommand \Eprint [0]{\href }%
\providecommand \doibase [0]{http://dx.doi.org/}%
\providecommand \selectlanguage [0]{\@gobble}%
\providecommand \bibinfo  [0]{\@secondoftwo}%
\providecommand \bibfield  [0]{\@secondoftwo}%
\providecommand \translation [1]{[#1]}%
\providecommand \BibitemOpen [0]{}%
\providecommand \bibitemStop [0]{}%
\providecommand \bibitemNoStop [0]{.\EOS\space}%
\providecommand \EOS [0]{\spacefactor3000\relax}%
\providecommand \BibitemShut  [1]{\csname bibitem#1\endcsname}%
\let\auto@bib@innerbib\@empty
\bibitem [{\citenamefont {Eisert}\ \emph {et~al.}(2015)\citenamefont {Eisert}, \citenamefont {Friesdorf},\ and\ \citenamefont {Gogolin}}]{eisert2015quantum}%
  \BibitemOpen
  \bibfield  {author} {\bibinfo {author} {\bibfnamefont {J.}~\bibnamefont {Eisert}}, \bibinfo {author} {\bibfnamefont {M.}~\bibnamefont {Friesdorf}}, \ and\ \bibinfo {author} {\bibfnamefont {C.}~\bibnamefont {Gogolin}},\ }\href@noop {} {\bibfield  {journal} {\bibinfo  {journal} {Nature Physics}\ }\textbf {\bibinfo {volume} {11}},\ \bibinfo {pages} {124} (\bibinfo {year} {2015})}\BibitemShut {NoStop}%
\bibitem [{\citenamefont {Gring}\ \emph {et~al.}(2012)\citenamefont {Gring}, \citenamefont {Kuhnert}, \citenamefont {Langen}, \citenamefont {Kitagawa}, \citenamefont {Rauer}, \citenamefont {Schreitl}, \citenamefont {Mazets}, \citenamefont {Smith}, \citenamefont {Demler},\ and\ \citenamefont {Schmiedmayer}}]{gring2012relaxation}%
  \BibitemOpen
  \bibfield  {author} {\bibinfo {author} {\bibfnamefont {M.}~\bibnamefont {Gring}}, \bibinfo {author} {\bibfnamefont {M.}~\bibnamefont {Kuhnert}}, \bibinfo {author} {\bibfnamefont {T.}~\bibnamefont {Langen}}, \bibinfo {author} {\bibfnamefont {T.}~\bibnamefont {Kitagawa}}, \bibinfo {author} {\bibfnamefont {B.}~\bibnamefont {Rauer}}, \bibinfo {author} {\bibfnamefont {M.}~\bibnamefont {Schreitl}}, \bibinfo {author} {\bibfnamefont {I.}~\bibnamefont {Mazets}}, \bibinfo {author} {\bibfnamefont {D.~A.}\ \bibnamefont {Smith}}, \bibinfo {author} {\bibfnamefont {E.}~\bibnamefont {Demler}}, \ and\ \bibinfo {author} {\bibfnamefont {J.}~\bibnamefont {Schmiedmayer}},\ }\href@noop {} {\bibfield  {journal} {\bibinfo  {journal} {Science}\ }\textbf {\bibinfo {volume} {337}},\ \bibinfo {pages} {1318} (\bibinfo {year} {2012})}\BibitemShut {NoStop}%
\bibitem [{\citenamefont {Simenel}\ \emph {et~al.}(2020)\citenamefont {Simenel}, \citenamefont {Godbey},\ and\ \citenamefont {Umar}}]{simenel2020timescales}%
  \BibitemOpen
  \bibfield  {author} {\bibinfo {author} {\bibfnamefont {C.}~\bibnamefont {Simenel}}, \bibinfo {author} {\bibfnamefont {K.}~\bibnamefont {Godbey}}, \ and\ \bibinfo {author} {\bibfnamefont {A.~S.}\ \bibnamefont {Umar}},\ }\href@noop {} {\bibfield  {journal} {\bibinfo  {journal} {Physical Review Letters}\ }\textbf {\bibinfo {volume} {124}},\ \bibinfo {pages} {212504} (\bibinfo {year} {2020})}\BibitemShut {NoStop}%
\bibitem [{\citenamefont {Schroder}\ and\ \citenamefont {Huizenga}(1977)}]{schroderDampedHeavyIonCollisions1977}%
  \BibitemOpen
  \bibfield  {author} {\bibinfo {author} {\bibfnamefont {W.~U.}\ \bibnamefont {Schroder}}\ and\ \bibinfo {author} {\bibfnamefont {J.~R.}\ \bibnamefont {Huizenga}},\ }\href@noop {} {\bibfield  {journal} {\bibinfo  {journal} {Annual Review of Nuclear Science}\ }\textbf {\bibinfo {volume} {27}},\ \bibinfo {pages} {465} (\bibinfo {year} {1977})}\BibitemShut {NoStop}%
\bibitem [{\citenamefont {Huizenga}\ \emph {et~al.}(1976)\citenamefont {Huizenga}, \citenamefont {Birkelund}, \citenamefont {Schr{\"o}der}, \citenamefont {Wolf},\ and\ \citenamefont {Viola}}]{huizenga1976energy}%
  \BibitemOpen
  \bibfield  {author} {\bibinfo {author} {\bibfnamefont {J.~R.}\ \bibnamefont {Huizenga}}, \bibinfo {author} {\bibfnamefont {J.~R.}\ \bibnamefont {Birkelund}}, \bibinfo {author} {\bibfnamefont {W.~U.}\ \bibnamefont {Schr{\"o}der}}, \bibinfo {author} {\bibfnamefont {K.~L.}\ \bibnamefont {Wolf}}, \ and\ \bibinfo {author} {\bibfnamefont {V.~E.}\ \bibnamefont {Viola}},\ }\href@noop {} {\bibfield  {journal} {\bibinfo  {journal} {Physical Review Letters}\ }\textbf {\bibinfo {volume} {37}},\ \bibinfo {pages} {885} (\bibinfo {year} {1976})}\BibitemShut {NoStop}%
\bibitem [{\citenamefont {Williams}\ \emph {et~al.}(2018)\citenamefont {Williams}, \citenamefont {Sekizawa}, \citenamefont {Hinde}, \citenamefont {Simenel}, \citenamefont {Dasgupta}, \citenamefont {Carter}, \citenamefont {Cook}, \citenamefont {Jeung}, \citenamefont {McNeil}, \citenamefont {Palshetkar}, \citenamefont {Rafferty}, \citenamefont {Ramachandran},\ and\ \citenamefont {Wakhle}}]{williams2018exploring}%
  \BibitemOpen
  \bibfield  {author} {\bibinfo {author} {\bibfnamefont {E.}~\bibnamefont {Williams}}, \bibinfo {author} {\bibfnamefont {K.}~\bibnamefont {Sekizawa}}, \bibinfo {author} {\bibfnamefont {D.~J.}\ \bibnamefont {Hinde}}, \bibinfo {author} {\bibfnamefont {C.}~\bibnamefont {Simenel}}, \bibinfo {author} {\bibfnamefont {M.}~\bibnamefont {Dasgupta}}, \bibinfo {author} {\bibfnamefont {I.~P.}\ \bibnamefont {Carter}}, \bibinfo {author} {\bibfnamefont {K.~J.}\ \bibnamefont {Cook}}, \bibinfo {author} {\bibfnamefont {D.~Y.}\ \bibnamefont {Jeung}}, \bibinfo {author} {\bibfnamefont {S.~D.}\ \bibnamefont {McNeil}}, \bibinfo {author} {\bibfnamefont {C.~S.}\ \bibnamefont {Palshetkar}}, \bibinfo {author} {\bibfnamefont {D.~C.}\ \bibnamefont {Rafferty}}, \bibinfo {author} {\bibfnamefont {K.}~\bibnamefont {Ramachandran}}, \ and\ \bibinfo {author} {\bibfnamefont {A.}~\bibnamefont {Wakhle}},\ }\href@noop {} {\bibfield  {journal} {\bibinfo  {journal} {Physical Review Letters}\ }\textbf {\bibinfo {volume} {120}},\ \bibinfo {pages} {022501} (\bibinfo {year} {2018})}\BibitemShut {NoStop}%
\bibitem [{\citenamefont {Jiang}\ \emph {et~al.}(2013)\citenamefont {Jiang}, \citenamefont {Yan},\ and\ \citenamefont {Maruhn}}]{jiang2013dynamics}%
  \BibitemOpen
  \bibfield  {author} {\bibinfo {author} {\bibfnamefont {X.}~\bibnamefont {Jiang}}, \bibinfo {author} {\bibfnamefont {S.}~\bibnamefont {Yan}}, \ and\ \bibinfo {author} {\bibfnamefont {J.~A.}\ \bibnamefont {Maruhn}},\ }\href@noop {} {\bibfield  {journal} {\bibinfo  {journal} {Physical Review C}\ }\textbf {\bibinfo {volume} {88}},\ \bibinfo {pages} {044611} (\bibinfo {year} {2013})}\BibitemShut {NoStop}%
\bibitem [{\citenamefont {Feldmeier}\ and\ \citenamefont {Spangenberger}(1984)}]{feldmeier1984particle}%
  \BibitemOpen
  \bibfield  {author} {\bibinfo {author} {\bibfnamefont {H.}~\bibnamefont {Feldmeier}}\ and\ \bibinfo {author} {\bibfnamefont {H.}~\bibnamefont {Spangenberger}},\ }\href@noop {} {\bibfield  {journal} {\bibinfo  {journal} {Nuclear Physics A}\ }\textbf {\bibinfo {volume} {428}},\ \bibinfo {pages} {223} (\bibinfo {year} {1984})}\BibitemShut {NoStop}%
\bibitem [{\citenamefont {Schr{\"o}der}\ \emph {et~al.}(1978)\citenamefont {Schr{\"o}der}, \citenamefont {Birkelund}, \citenamefont {Huizenga}, \citenamefont {Wolf},\ and\ \citenamefont {Viola}}]{schroder1978mechanisms}%
  \BibitemOpen
  \bibfield  {author} {\bibinfo {author} {\bibfnamefont {W.}~\bibnamefont {Schr{\"o}der}}, \bibinfo {author} {\bibfnamefont {J.}~\bibnamefont {Birkelund}}, \bibinfo {author} {\bibfnamefont {J.}~\bibnamefont {Huizenga}}, \bibinfo {author} {\bibfnamefont {K.}~\bibnamefont {Wolf}}, \ and\ \bibinfo {author} {\bibfnamefont {V.}~\bibnamefont {Viola}},\ }\href@noop {} {\bibfield  {journal} {\bibinfo  {journal} {Physics Reports}\ }\textbf {\bibinfo {volume} {45}},\ \bibinfo {pages} {301} (\bibinfo {year} {1978})}\BibitemShut {NoStop}%
\bibitem [{\citenamefont {Wilcke}\ \emph {et~al.}(1980)\citenamefont {Wilcke}, \citenamefont {Birkelund}, \citenamefont {Hoover}, \citenamefont {Huizenga}, \citenamefont {Schr{\"o}der}, \citenamefont {Viola}, \citenamefont {Wolf},\ and\ \citenamefont {Mignerey}}]{wilcke1980bombardingenergy}%
  \BibitemOpen
  \bibfield  {author} {\bibinfo {author} {\bibfnamefont {W.~W.}\ \bibnamefont {Wilcke}}, \bibinfo {author} {\bibfnamefont {J.~R.}\ \bibnamefont {Birkelund}}, \bibinfo {author} {\bibfnamefont {A.~D.}\ \bibnamefont {Hoover}}, \bibinfo {author} {\bibfnamefont {J.~R.}\ \bibnamefont {Huizenga}}, \bibinfo {author} {\bibfnamefont {W.~U.}\ \bibnamefont {Schr{\"o}der}}, \bibinfo {author} {\bibfnamefont {V.~E.}\ \bibnamefont {Viola}}, \bibinfo {author} {\bibfnamefont {K.~L.}\ \bibnamefont {Wolf}}, \ and\ \bibinfo {author} {\bibfnamefont {A.~C.}\ \bibnamefont {Mignerey}},\ }\href@noop {} {\bibfield  {journal} {\bibinfo  {journal} {Physical Review C}\ }\textbf {\bibinfo {volume} {22}},\ \bibinfo {pages} {128} (\bibinfo {year} {1980})}\BibitemShut {NoStop}%
\bibitem [{\citenamefont {Wollersheim}\ \emph {et~al.}(1981)\citenamefont {Wollersheim}, \citenamefont {Wilcke}, \citenamefont {Birkelund}, \citenamefont {Huizenga}, \citenamefont {Schr{\"o}der}, \citenamefont {Freiesleben},\ and\ \citenamefont {Hilscher}}]{wollersheim1981bi}%
  \BibitemOpen
  \bibfield  {author} {\bibinfo {author} {\bibfnamefont {H.~J.}\ \bibnamefont {Wollersheim}}, \bibinfo {author} {\bibfnamefont {W.~W.}\ \bibnamefont {Wilcke}}, \bibinfo {author} {\bibfnamefont {J.~R.}\ \bibnamefont {Birkelund}}, \bibinfo {author} {\bibfnamefont {J.~R.}\ \bibnamefont {Huizenga}}, \bibinfo {author} {\bibfnamefont {W.~U.}\ \bibnamefont {Schr{\"o}der}}, \bibinfo {author} {\bibfnamefont {H.}~\bibnamefont {Freiesleben}}, \ and\ \bibinfo {author} {\bibfnamefont {D.}~\bibnamefont {Hilscher}},\ }\href@noop {} {\bibfield  {journal} {\bibinfo  {journal} {Physical Review C}\ }\textbf {\bibinfo {volume} {24}},\ \bibinfo {pages} {2114} (\bibinfo {year} {1981})}\BibitemShut {NoStop}%
\bibitem [{\citenamefont {Liao}\ \emph {et~al.}(2023)\citenamefont {Liao}, \citenamefont {Zhu}, \citenamefont {Su},\ and\ \citenamefont {Li}}]{liao2023dynamics}%
  \BibitemOpen
  \bibfield  {author} {\bibinfo {author} {\bibfnamefont {Z.}~\bibnamefont {Liao}}, \bibinfo {author} {\bibfnamefont {L.}~\bibnamefont {Zhu}}, \bibinfo {author} {\bibfnamefont {J.}~\bibnamefont {Su}}, \ and\ \bibinfo {author} {\bibfnamefont {C.}~\bibnamefont {Li}},\ }\href@noop {} {\bibfield  {journal} {\bibinfo  {journal} {Physical Review C}\ }\textbf {\bibinfo {volume} {107}},\ \bibinfo {pages} {014614} (\bibinfo {year} {2023})}\BibitemShut {NoStop}%
\bibitem [{\citenamefont {Jedele}\ \emph {et~al.}(2017)\citenamefont {Jedele}, \citenamefont {McIntosh}, \citenamefont {Hagel}, \citenamefont {Huang}, \citenamefont {Heilborn}, \citenamefont {Kohley}, \citenamefont {May}, \citenamefont {McCleskey}, \citenamefont {Youngs}, \citenamefont {Zarrella},\ and\ \citenamefont {Yennello}}]{jedele2017characterizing}%
  \BibitemOpen
  \bibfield  {author} {\bibinfo {author} {\bibfnamefont {A.}~\bibnamefont {Jedele}}, \bibinfo {author} {\bibfnamefont {A.~B.}\ \bibnamefont {McIntosh}}, \bibinfo {author} {\bibfnamefont {K.}~\bibnamefont {Hagel}}, \bibinfo {author} {\bibfnamefont {M.}~\bibnamefont {Huang}}, \bibinfo {author} {\bibfnamefont {L.}~\bibnamefont {Heilborn}}, \bibinfo {author} {\bibfnamefont {Z.}~\bibnamefont {Kohley}}, \bibinfo {author} {\bibfnamefont {L.~W.}\ \bibnamefont {May}}, \bibinfo {author} {\bibfnamefont {E.}~\bibnamefont {McCleskey}}, \bibinfo {author} {\bibfnamefont {M.}~\bibnamefont {Youngs}}, \bibinfo {author} {\bibfnamefont {A.}~\bibnamefont {Zarrella}}, \ and\ \bibinfo {author} {\bibfnamefont {S.~J.}\ \bibnamefont {Yennello}},\ }\href@noop {} {\bibfield  {journal} {\bibinfo  {journal} {Physical Review Letters}\ }\textbf {\bibinfo {volume} {118}},\ \bibinfo {pages} {062501} (\bibinfo {year} {2017})}\BibitemShut {NoStop}%
\bibitem [{\citenamefont {Kr{\'o}las}\ \emph {et~al.}(2010)\citenamefont {Kr{\'o}las}, \citenamefont {Broda}, \citenamefont {Fornal}, \citenamefont {Paw{\l}at}, \citenamefont {Wrzesi{\'n}ski}, \citenamefont {Bazzacco}, \citenamefont {De~Angelis}, \citenamefont {Lunardi}, \citenamefont {Menegazzo}, \citenamefont {Napoli},\ and\ \citenamefont {Rossi~Alvarez}}]{krolas2010dynamical}%
  \BibitemOpen
  \bibfield  {author} {\bibinfo {author} {\bibfnamefont {W.}~\bibnamefont {Kr{\'o}las}}, \bibinfo {author} {\bibfnamefont {R.}~\bibnamefont {Broda}}, \bibinfo {author} {\bibfnamefont {B.}~\bibnamefont {Fornal}}, \bibinfo {author} {\bibfnamefont {T.}~\bibnamefont {Paw{\l}at}}, \bibinfo {author} {\bibfnamefont {J.}~\bibnamefont {Wrzesi{\'n}ski}}, \bibinfo {author} {\bibfnamefont {D.}~\bibnamefont {Bazzacco}}, \bibinfo {author} {\bibfnamefont {G.}~\bibnamefont {De~Angelis}}, \bibinfo {author} {\bibfnamefont {S.}~\bibnamefont {Lunardi}}, \bibinfo {author} {\bibfnamefont {R.}~\bibnamefont {Menegazzo}}, \bibinfo {author} {\bibfnamefont {D.}~\bibnamefont {Napoli}}, \ and\ \bibinfo {author} {\bibfnamefont {C.}~\bibnamefont {Rossi~Alvarez}},\ }\href@noop {} {\bibfield  {journal} {\bibinfo  {journal} {Nuclear Physics A}\ }\textbf {\bibinfo {volume} {832}},\ \bibinfo {pages} {170} (\bibinfo {year} {2010})}\BibitemShut {NoStop}%
\bibitem [{\citenamefont {Kratz}\ \emph {et~al.}(1977)\citenamefont {Kratz}, \citenamefont {Ahrens}, \citenamefont {B{\"o}gl}, \citenamefont {Br{\"u}chle}, \citenamefont {Franz}, \citenamefont {Sch{\"a}del}, \citenamefont {Warnecke}, \citenamefont {Wirth}, \citenamefont {Klein},\ and\ \citenamefont {Weis}}]{kratz1977chargeasymmetry}%
  \BibitemOpen
  \bibfield  {author} {\bibinfo {author} {\bibfnamefont {J.~V.}\ \bibnamefont {Kratz}}, \bibinfo {author} {\bibfnamefont {H.}~\bibnamefont {Ahrens}}, \bibinfo {author} {\bibfnamefont {W.}~\bibnamefont {B{\"o}gl}}, \bibinfo {author} {\bibfnamefont {W.}~\bibnamefont {Br{\"u}chle}}, \bibinfo {author} {\bibfnamefont {G.}~\bibnamefont {Franz}}, \bibinfo {author} {\bibfnamefont {M.}~\bibnamefont {Sch{\"a}del}}, \bibinfo {author} {\bibfnamefont {I.}~\bibnamefont {Warnecke}}, \bibinfo {author} {\bibfnamefont {G.}~\bibnamefont {Wirth}}, \bibinfo {author} {\bibfnamefont {G.}~\bibnamefont {Klein}}, \ and\ \bibinfo {author} {\bibfnamefont {M.}~\bibnamefont {Weis}},\ }\href@noop {} {\bibfield  {journal} {\bibinfo  {journal} {Physical Review Letters}\ }\textbf {\bibinfo {volume} {39}},\ \bibinfo {pages} {984} (\bibinfo {year} {1977})}\BibitemShut {NoStop}%
\bibitem [{\citenamefont {Hernandez}\ \emph {et~al.}(1981)\citenamefont {Hernandez}, \citenamefont {Myers}, \citenamefont {Randrup},\ and\ \citenamefont {Remaud}}]{hernandez1981quantal}%
  \BibitemOpen
  \bibfield  {author} {\bibinfo {author} {\bibfnamefont {E.}~\bibnamefont {Hernandez}}, \bibinfo {author} {\bibfnamefont {W.}~\bibnamefont {Myers}}, \bibinfo {author} {\bibfnamefont {J.}~\bibnamefont {Randrup}}, \ and\ \bibinfo {author} {\bibfnamefont {B.}~\bibnamefont {Remaud}},\ }\href@noop {} {\bibfield  {journal} {\bibinfo  {journal} {Nuclear Physics A}\ }\textbf {\bibinfo {volume} {361}},\ \bibinfo {pages} {483} (\bibinfo {year} {1981})}\BibitemShut {NoStop}%
\bibitem [{\citenamefont {Mathews}\ \emph {et~al.}(1982)\citenamefont {Mathews}, \citenamefont {Moulton}, \citenamefont {Wozniak}, \citenamefont {Cauvin}, \citenamefont {Schmitt}, \citenamefont {Sventek},\ and\ \citenamefont {Moretto}}]{mathewsNe20Induced1982}%
  \BibitemOpen
  \bibfield  {author} {\bibinfo {author} {\bibfnamefont {G.~J.}\ \bibnamefont {Mathews}}, \bibinfo {author} {\bibfnamefont {J.~B.}\ \bibnamefont {Moulton}}, \bibinfo {author} {\bibfnamefont {G.~J.}\ \bibnamefont {Wozniak}}, \bibinfo {author} {\bibfnamefont {B.}~\bibnamefont {Cauvin}}, \bibinfo {author} {\bibfnamefont {R.~P.}\ \bibnamefont {Schmitt}}, \bibinfo {author} {\bibfnamefont {J.~S.}\ \bibnamefont {Sventek}}, \ and\ \bibinfo {author} {\bibfnamefont {L.~G.}\ \bibnamefont {Moretto}},\ }\href@noop {} {\bibfield  {journal} {\bibinfo  {journal} {Physical Review C}\ }\textbf {\bibinfo {volume} {25}},\ \bibinfo {pages} {300} (\bibinfo {year} {1982})}\BibitemShut {NoStop}%
\bibitem [{\citenamefont {Freiesleben}\ and\ \citenamefont {Kratz}(1984)}]{freiesleben1984nzequilibration}%
  \BibitemOpen
  \bibfield  {author} {\bibinfo {author} {\bibfnamefont {H.}~\bibnamefont {Freiesleben}}\ and\ \bibinfo {author} {\bibfnamefont {J.}~\bibnamefont {Kratz}},\ }\href@noop {} {\bibfield  {journal} {\bibinfo  {journal} {Physics Reports}\ }\textbf {\bibinfo {volume} {106}},\ \bibinfo {pages} {1} (\bibinfo {year} {1984})}\BibitemShut {NoStop}%
\bibitem [{\citenamefont {Moretto}\ \emph {et~al.}(1979)\citenamefont {Moretto}, \citenamefont {Sventek},\ and\ \citenamefont {Mantzouranis}}]{moretto1979giant}%
  \BibitemOpen
  \bibfield  {author} {\bibinfo {author} {\bibfnamefont {L.~G.}\ \bibnamefont {Moretto}}, \bibinfo {author} {\bibfnamefont {J.}~\bibnamefont {Sventek}}, \ and\ \bibinfo {author} {\bibfnamefont {G.}~\bibnamefont {Mantzouranis}},\ }\href@noop {} {\bibfield  {journal} {\bibinfo  {journal} {Physical Review Letters}\ }\textbf {\bibinfo {volume} {42}},\ \bibinfo {pages} {563} (\bibinfo {year} {1979})}\BibitemShut {NoStop}%
\bibitem [{\citenamefont {Brosa}\ and\ \citenamefont {Krappe}(1978)}]{brosa1978time}%
  \BibitemOpen
  \bibfield  {author} {\bibinfo {author} {\bibfnamefont {U.}~\bibnamefont {Brosa}}\ and\ \bibinfo {author} {\bibfnamefont {H.~J.}\ \bibnamefont {Krappe}},\ }\href@noop {} {\bibfield  {journal} {\bibinfo  {journal} {Zeitschrift f{\"u}r Physik A Atoms and Nuclei}\ }\textbf {\bibinfo {volume} {284}},\ \bibinfo {pages} {65} (\bibinfo {year} {1978})}\BibitemShut {NoStop}%
\bibitem [{\citenamefont {Berlanger}\ \emph {et~al.}(1979)\citenamefont {Berlanger}, \citenamefont {Gobbi}, \citenamefont {Hanappe}, \citenamefont {Lynen}, \citenamefont {Ng{\^o}}, \citenamefont {Olmi}, \citenamefont {Sann}, \citenamefont {Stelzer}, \citenamefont {Richel},\ and\ \citenamefont {Rivet}}]{berlanger1979studya}%
  \BibitemOpen
  \bibfield  {author} {\bibinfo {author} {\bibfnamefont {M.}~\bibnamefont {Berlanger}}, \bibinfo {author} {\bibfnamefont {A.}~\bibnamefont {Gobbi}}, \bibinfo {author} {\bibfnamefont {F.}~\bibnamefont {Hanappe}}, \bibinfo {author} {\bibfnamefont {U.}~\bibnamefont {Lynen}}, \bibinfo {author} {\bibfnamefont {C.}~\bibnamefont {Ng{\^o}}}, \bibinfo {author} {\bibfnamefont {A.}~\bibnamefont {Olmi}}, \bibinfo {author} {\bibfnamefont {H.}~\bibnamefont {Sann}}, \bibinfo {author} {\bibfnamefont {H.}~\bibnamefont {Stelzer}}, \bibinfo {author} {\bibfnamefont {H.}~\bibnamefont {Richel}}, \ and\ \bibinfo {author} {\bibfnamefont {M.}~\bibnamefont {Rivet}},\ }\href@noop {} {\bibfield  {journal} {\bibinfo  {journal} {Zeitschrift f{\"u}r Physik A Atoms and Nuclei}\ }\textbf {\bibinfo {volume} {291}},\ \bibinfo {pages} {133} (\bibinfo {year} {1979})}\BibitemShut {NoStop}%
\bibitem [{\citenamefont {Hofmann}\ \emph {et~al.}(1979)\citenamefont {Hofmann}, \citenamefont {Gr{\'e}goire}, \citenamefont {Lucas},\ and\ \citenamefont {Ng{\^o}}}]{hofmann1979theoretical}%
  \BibitemOpen
  \bibfield  {author} {\bibinfo {author} {\bibfnamefont {H.}~\bibnamefont {Hofmann}}, \bibinfo {author} {\bibfnamefont {C.}~\bibnamefont {Gr{\'e}goire}}, \bibinfo {author} {\bibfnamefont {R.}~\bibnamefont {Lucas}}, \ and\ \bibinfo {author} {\bibfnamefont {C.}~\bibnamefont {Ng{\^o}}},\ }\href@noop {} {\bibfield  {journal} {\bibinfo  {journal} {Zeitschrift f{\"u}r Physik A Atoms and Nuclei}\ }\textbf {\bibinfo {volume} {293}},\ \bibinfo {pages} {229} (\bibinfo {year} {1979})}\BibitemShut {NoStop}%
\bibitem [{\citenamefont {Wu}\ \emph {et~al.}(2010)\citenamefont {Wu}, \citenamefont {Tian}, \citenamefont {Ma}, \citenamefont {Cai}, \citenamefont {Chen}, \citenamefont {Fang}, \citenamefont {Guo},\ and\ \citenamefont {Wang}}]{wuDynamicalDipoleRadiation2010}%
  \BibitemOpen
  \bibfield  {author} {\bibinfo {author} {\bibfnamefont {H.~L.}\ \bibnamefont {Wu}}, \bibinfo {author} {\bibfnamefont {W.~D.}\ \bibnamefont {Tian}}, \bibinfo {author} {\bibfnamefont {Y.~G.}\ \bibnamefont {Ma}}, \bibinfo {author} {\bibfnamefont {X.~Z.}\ \bibnamefont {Cai}}, \bibinfo {author} {\bibfnamefont {J.~G.}\ \bibnamefont {Chen}}, \bibinfo {author} {\bibfnamefont {D.~Q.}\ \bibnamefont {Fang}}, \bibinfo {author} {\bibfnamefont {W.}~\bibnamefont {Guo}}, \ and\ \bibinfo {author} {\bibfnamefont {H.~W.}\ \bibnamefont {Wang}},\ }\href@noop {} {\bibfield  {journal} {\bibinfo  {journal} {Physical Review C}\ }\textbf {\bibinfo {volume} {81}},\ \bibinfo {pages} {047602} (\bibinfo {year} {2010})}\BibitemShut {NoStop}%
\bibitem [{\citenamefont {Iwata}\ \emph {et~al.}(2010)\citenamefont {Iwata}, \citenamefont {Otsuka}, \citenamefont {Maruhn},\ and\ \citenamefont {Itagaki}}]{iwata2010suppression}%
  \BibitemOpen
  \bibfield  {author} {\bibinfo {author} {\bibfnamefont {Y.}~\bibnamefont {Iwata}}, \bibinfo {author} {\bibfnamefont {T.}~\bibnamefont {Otsuka}}, \bibinfo {author} {\bibfnamefont {J.~A.}\ \bibnamefont {Maruhn}}, \ and\ \bibinfo {author} {\bibfnamefont {N.}~\bibnamefont {Itagaki}},\ }\href@noop {} {\bibfield  {journal} {\bibinfo  {journal} {Physical Review Letters}\ }\textbf {\bibinfo {volume} {104}},\ \bibinfo {pages} {252501} (\bibinfo {year} {2010})}\BibitemShut {NoStop}%
\bibitem [{\citenamefont {Zhang}\ \emph {et~al.}(1999)\citenamefont {Zhang}, \citenamefont {Chen}, \citenamefont {Ming},\ and\ \citenamefont {Zhu}}]{zhang1999isospin}%
  \BibitemOpen
  \bibfield  {author} {\bibinfo {author} {\bibfnamefont {F.-S.}\ \bibnamefont {Zhang}}, \bibinfo {author} {\bibfnamefont {L.-W.}\ \bibnamefont {Chen}}, \bibinfo {author} {\bibfnamefont {Z.-Y.}\ \bibnamefont {Ming}}, \ and\ \bibinfo {author} {\bibfnamefont {Z.-Y.}\ \bibnamefont {Zhu}},\ }\href@noop {} {\bibfield  {journal} {\bibinfo  {journal} {Physical Review C}\ }\textbf {\bibinfo {volume} {60}},\ \bibinfo {pages} {064604} (\bibinfo {year} {1999})}\BibitemShut {NoStop}%
\bibitem [{\citenamefont {Chen}\ \emph {et~al.}(1999)\citenamefont {Chen}, \citenamefont {Zhang}, \citenamefont {Jin},\ and\ \citenamefont {Zhu}}]{chen1999isospin}%
  \BibitemOpen
  \bibfield  {author} {\bibinfo {author} {\bibfnamefont {L.-W.}\ \bibnamefont {Chen}}, \bibinfo {author} {\bibfnamefont {F.-S.}\ \bibnamefont {Zhang}}, \bibinfo {author} {\bibfnamefont {G.-M.}\ \bibnamefont {Jin}}, \ and\ \bibinfo {author} {\bibfnamefont {Z.-Y.}\ \bibnamefont {Zhu}},\ }\href@noop {} {\bibfield  {journal} {\bibinfo  {journal} {Physics Letters B}\ }\textbf {\bibinfo {volume} {459}},\ \bibinfo {pages} {21} (\bibinfo {year} {1999})}\BibitemShut {NoStop}%
\bibitem [{\citenamefont {Chen}\ \emph {et~al.}(2000)\citenamefont {Chen}, \citenamefont {Zhang},\ and\ \citenamefont {Zhu}}]{chen2000isospin}%
  \BibitemOpen
  \bibfield  {author} {\bibinfo {author} {\bibfnamefont {L.-W.}\ \bibnamefont {Chen}}, \bibinfo {author} {\bibfnamefont {F.-S.}\ \bibnamefont {Zhang}}, \ and\ \bibinfo {author} {\bibfnamefont {Z.-Y.}\ \bibnamefont {Zhu}},\ }\href@noop {} {\bibfield  {journal} {\bibinfo  {journal} {Physical Review C}\ }\textbf {\bibinfo {volume} {61}},\ \bibinfo {pages} {067601} (\bibinfo {year} {2000})}\BibitemShut {NoStop}%
\bibitem [{\citenamefont {Aichelin}(1991)}]{aichelin1991quantum}%
  \BibitemOpen
  \bibfield  {author} {\bibinfo {author} {\bibfnamefont {J.}~\bibnamefont {Aichelin}},\ }\href@noop {} {\bibfield  {journal} {\bibinfo  {journal} {Physics Reports}\ }\textbf {\bibinfo {volume} {202}},\ \bibinfo {pages} {233} (\bibinfo {year} {1991})}\BibitemShut {NoStop}%
\bibitem [{\citenamefont {Slater}(1951)}]{slater1951simplification}%
  \BibitemOpen
  \bibfield  {author} {\bibinfo {author} {\bibfnamefont {J.~C.}\ \bibnamefont {Slater}},\ }\href@noop {} {\bibfield  {journal} {\bibinfo  {journal} {Physical Review}\ }\textbf {\bibinfo {volume} {81}},\ \bibinfo {pages} {385} (\bibinfo {year} {1951})}\BibitemShut {NoStop}%
\bibitem [{\citenamefont {Wang}\ \emph {et~al.}(2002)\citenamefont {Wang}, \citenamefont {Li},\ and\ \citenamefont {Wu}}]{wang2002improved}%
  \BibitemOpen
  \bibfield  {author} {\bibinfo {author} {\bibfnamefont {N.}~\bibnamefont {Wang}}, \bibinfo {author} {\bibfnamefont {Z.}~\bibnamefont {Li}}, \ and\ \bibinfo {author} {\bibfnamefont {X.}~\bibnamefont {Wu}},\ }\href@noop {} {\bibfield  {journal} {\bibinfo  {journal} {Physical Review C}\ }\textbf {\bibinfo {volume} {65}},\ \bibinfo {pages} {064608} (\bibinfo {year} {2002})}\BibitemShut {NoStop}%
\bibitem [{\citenamefont {Papa}\ \emph {et~al.}(2005)\citenamefont {Papa}, \citenamefont {Giuliani},\ and\ \citenamefont {Bonasera}}]{papa2005constrained}%
  \BibitemOpen
  \bibfield  {author} {\bibinfo {author} {\bibfnamefont {M.}~\bibnamefont {Papa}}, \bibinfo {author} {\bibfnamefont {G.}~\bibnamefont {Giuliani}}, \ and\ \bibinfo {author} {\bibfnamefont {A.}~\bibnamefont {Bonasera}},\ }\href@noop {} {\bibfield  {journal} {\bibinfo  {journal} {Journal of Computational Physics}\ }\textbf {\bibinfo {volume} {208}},\ \bibinfo {pages} {403} (\bibinfo {year} {2005})}\BibitemShut {NoStop}%
\bibitem [{\citenamefont {Maruyama}\ \emph {et~al.}(1990)\citenamefont {Maruyama}, \citenamefont {Ohnishi},\ and\ \citenamefont {Horiuchi}}]{maruyama1990quantum}%
  \BibitemOpen
  \bibfield  {author} {\bibinfo {author} {\bibfnamefont {T.}~\bibnamefont {Maruyama}}, \bibinfo {author} {\bibfnamefont {A.}~\bibnamefont {Ohnishi}}, \ and\ \bibinfo {author} {\bibfnamefont {H.}~\bibnamefont {Horiuchi}},\ }\href@noop {} {\bibfield  {journal} {\bibinfo  {journal} {Physical Review C}\ }\textbf {\bibinfo {volume} {42}},\ \bibinfo {pages} {386} (\bibinfo {year} {1990})}\BibitemShut {NoStop}%
\bibitem [{\citenamefont {Zhao}\ \emph {et~al.}(2009)\citenamefont {Zhao}, \citenamefont {Li}, \citenamefont {Wu},\ and\ \citenamefont {Zhao}}]{zhao2009mass}%
  \BibitemOpen
  \bibfield  {author} {\bibinfo {author} {\bibfnamefont {K.}~\bibnamefont {Zhao}}, \bibinfo {author} {\bibfnamefont {Z.}~\bibnamefont {Li}}, \bibinfo {author} {\bibfnamefont {X.}~\bibnamefont {Wu}}, \ and\ \bibinfo {author} {\bibfnamefont {Z.}~\bibnamefont {Zhao}},\ }\href@noop {} {\bibfield  {journal} {\bibinfo  {journal} {Physical Review C}\ }\textbf {\bibinfo {volume} {79}},\ \bibinfo {pages} {024614} (\bibinfo {year} {2009})}\BibitemShut {NoStop}%
\bibitem [{\citenamefont {Li}\ \emph {et~al.}(2016)\citenamefont {Li}, \citenamefont {Zhang}, \citenamefont {Li}, \citenamefont {Zhu}, \citenamefont {Tian}, \citenamefont {Wang},\ and\ \citenamefont {Zhang}}]{li2016multinucleon}%
  \BibitemOpen
  \bibfield  {author} {\bibinfo {author} {\bibfnamefont {C.}~\bibnamefont {Li}}, \bibinfo {author} {\bibfnamefont {F.}~\bibnamefont {Zhang}}, \bibinfo {author} {\bibfnamefont {J.}~\bibnamefont {Li}}, \bibinfo {author} {\bibfnamefont {L.}~\bibnamefont {Zhu}}, \bibinfo {author} {\bibfnamefont {J.}~\bibnamefont {Tian}}, \bibinfo {author} {\bibfnamefont {N.}~\bibnamefont {Wang}}, \ and\ \bibinfo {author} {\bibfnamefont {F.-S.}\ \bibnamefont {Zhang}},\ }\href@noop {} {\bibfield  {journal} {\bibinfo  {journal} {Physical Review C}\ }\textbf {\bibinfo {volume} {93}},\ \bibinfo {pages} {014618} (\bibinfo {year} {2016})}\BibitemShut {NoStop}%
\bibitem [{\citenamefont {Li}\ \emph {et~al.}(2020)\citenamefont {Li}, \citenamefont {Tian},\ and\ \citenamefont {Zhang}}]{li2020production}%
  \BibitemOpen
  \bibfield  {author} {\bibinfo {author} {\bibfnamefont {C.}~\bibnamefont {Li}}, \bibinfo {author} {\bibfnamefont {J.}~\bibnamefont {Tian}}, \ and\ \bibinfo {author} {\bibfnamefont {F.-S.}\ \bibnamefont {Zhang}},\ }\href@noop {} {\bibfield  {journal} {\bibinfo  {journal} {Physics Letters B}\ }\textbf {\bibinfo {volume} {809}},\ \bibinfo {pages} {135697} (\bibinfo {year} {2020})}\BibitemShut {NoStop}%
\bibitem [{\citenamefont {Li}\ \emph {et~al.}(2018)\citenamefont {Li}, \citenamefont {Wen}, \citenamefont {Li}, \citenamefont {Zhang}, \citenamefont {Li}, \citenamefont {Xu}, \citenamefont {Liu}, \citenamefont {Zhu},\ and\ \citenamefont {Zhang}}]{li2018production}%
  \BibitemOpen
  \bibfield  {author} {\bibinfo {author} {\bibfnamefont {C.}~\bibnamefont {Li}}, \bibinfo {author} {\bibfnamefont {P.}~\bibnamefont {Wen}}, \bibinfo {author} {\bibfnamefont {J.}~\bibnamefont {Li}}, \bibinfo {author} {\bibfnamefont {G.}~\bibnamefont {Zhang}}, \bibinfo {author} {\bibfnamefont {B.}~\bibnamefont {Li}}, \bibinfo {author} {\bibfnamefont {X.}~\bibnamefont {Xu}}, \bibinfo {author} {\bibfnamefont {Z.}~\bibnamefont {Liu}}, \bibinfo {author} {\bibfnamefont {S.}~\bibnamefont {Zhu}}, \ and\ \bibinfo {author} {\bibfnamefont {F.-S.}\ \bibnamefont {Zhang}},\ }\href@noop {} {\bibfield  {journal} {\bibinfo  {journal} {Physics Letters B}\ }\textbf {\bibinfo {volume} {776}},\ \bibinfo {pages} {278} (\bibinfo {year} {2018})}\BibitemShut {NoStop}%
\bibitem [{\citenamefont {Yilmaz}\ \emph {et~al.}(2011)\citenamefont {Yilmaz}, \citenamefont {Ayik}, \citenamefont {Lacroix},\ and\ \citenamefont {Washiyama}}]{yilmaz2011nucleon}%
  \BibitemOpen
  \bibfield  {author} {\bibinfo {author} {\bibfnamefont {B.}~\bibnamefont {Yilmaz}}, \bibinfo {author} {\bibfnamefont {S.}~\bibnamefont {Ayik}}, \bibinfo {author} {\bibfnamefont {D.}~\bibnamefont {Lacroix}}, \ and\ \bibinfo {author} {\bibfnamefont {K.}~\bibnamefont {Washiyama}},\ }\href@noop {} {\bibfield  {journal} {\bibinfo  {journal} {Physical Review C}\ }\textbf {\bibinfo {volume} {83}},\ \bibinfo {pages} {064615} (\bibinfo {year} {2011})}\BibitemShut {NoStop}%
\bibitem [{\citenamefont {Yilmaz}\ \emph {et~al.}(2014)\citenamefont {Yilmaz}, \citenamefont {Ayik}, \citenamefont {Lacroix},\ and\ \citenamefont {Yilmaz}}]{yilmaz2014nucleon}%
  \BibitemOpen
  \bibfield  {author} {\bibinfo {author} {\bibfnamefont {B.}~\bibnamefont {Yilmaz}}, \bibinfo {author} {\bibfnamefont {S.}~\bibnamefont {Ayik}}, \bibinfo {author} {\bibfnamefont {D.}~\bibnamefont {Lacroix}}, \ and\ \bibinfo {author} {\bibfnamefont {O.}~\bibnamefont {Yilmaz}},\ }\href@noop {} {\bibfield  {journal} {\bibinfo  {journal} {Physical Review C}\ }\textbf {\bibinfo {volume} {90}},\ \bibinfo {pages} {024613} (\bibinfo {year} {2014})}\BibitemShut {NoStop}%
\bibitem [{\citenamefont {Feng}\ \emph {et~al.}(2023)\citenamefont {Feng}, \citenamefont {Huang}, \citenamefont {Xiao}, \citenamefont {Lei}, \citenamefont {Zhu},\ and\ \citenamefont {Su}}]{feng2023contributions}%
  \BibitemOpen
  \bibfield  {author} {\bibinfo {author} {\bibfnamefont {Y.}~\bibnamefont {Feng}}, \bibinfo {author} {\bibfnamefont {Y.}~\bibnamefont {Huang}}, \bibinfo {author} {\bibfnamefont {E.}~\bibnamefont {Xiao}}, \bibinfo {author} {\bibfnamefont {X.}~\bibnamefont {Lei}}, \bibinfo {author} {\bibfnamefont {L.}~\bibnamefont {Zhu}}, \ and\ \bibinfo {author} {\bibfnamefont {J.}~\bibnamefont {Su}},\ }\href@noop {} {\bibfield  {journal} {\bibinfo  {journal} {Physical Review C}\ }\textbf {\bibinfo {volume} {107}},\ \bibinfo {pages} {044606} (\bibinfo {year} {2023})}\BibitemShut {NoStop}%
\bibitem [{\citenamefont {Feng}\ \emph {et~al.}(2024)\citenamefont {Feng}, \citenamefont {Liu}, \citenamefont {Huang}, \citenamefont {Gu}, \citenamefont {Xiao}, \citenamefont {Lei}, \citenamefont {Wang}, \citenamefont {Huang}, \citenamefont {Zhu},\ and\ \citenamefont {Su}}]{feng2024microscopic}%
  \BibitemOpen
  \bibfield  {author} {\bibinfo {author} {\bibfnamefont {Y.}~\bibnamefont {Feng}}, \bibinfo {author} {\bibfnamefont {H.}~\bibnamefont {Liu}}, \bibinfo {author} {\bibfnamefont {Y.}~\bibnamefont {Huang}}, \bibinfo {author} {\bibfnamefont {F.}~\bibnamefont {Gu}}, \bibinfo {author} {\bibfnamefont {E.}~\bibnamefont {Xiao}}, \bibinfo {author} {\bibfnamefont {X.}~\bibnamefont {Lei}}, \bibinfo {author} {\bibfnamefont {H.}~\bibnamefont {Wang}}, \bibinfo {author} {\bibfnamefont {J.}~\bibnamefont {Huang}}, \bibinfo {author} {\bibfnamefont {L.}~\bibnamefont {Zhu}}, \ and\ \bibinfo {author} {\bibfnamefont {J.}~\bibnamefont {Su}},\ }\href@noop {} {\bibfield  {journal} {\bibinfo  {journal} {Physical Review C}\ }\textbf {\bibinfo {volume} {109}},\ \bibinfo {pages} {054604} (\bibinfo {year} {2024})}\BibitemShut {NoStop}%
\bibitem [{\citenamefont {Umar}\ \emph {et~al.}(2017)\citenamefont {Umar}, \citenamefont {Simenel},\ and\ \citenamefont {Ye}}]{umar2017transport}%
  \BibitemOpen
  \bibfield  {author} {\bibinfo {author} {\bibfnamefont {A.~S.}\ \bibnamefont {Umar}}, \bibinfo {author} {\bibfnamefont {C.}~\bibnamefont {Simenel}}, \ and\ \bibinfo {author} {\bibfnamefont {W.}~\bibnamefont {Ye}},\ }\href@noop {} {\bibfield  {journal} {\bibinfo  {journal} {Physical Review C}\ }\textbf {\bibinfo {volume} {96}},\ \bibinfo {pages} {024625} (\bibinfo {year} {2017})}\BibitemShut {NoStop}%
\bibitem [{\citenamefont {Rehm}\ \emph {et~al.}(1979)\citenamefont {Rehm}, \citenamefont {Essel}, \citenamefont {Hartel}, \citenamefont {Kienle}, \citenamefont {K{\"o}rner}, \citenamefont {Segel}, \citenamefont {Sperr},\ and\ \citenamefont {Wagner}}]{rehm1979timea}%
  \BibitemOpen
  \bibfield  {author} {\bibinfo {author} {\bibfnamefont {K.~E.}\ \bibnamefont {Rehm}}, \bibinfo {author} {\bibfnamefont {H.}~\bibnamefont {Essel}}, \bibinfo {author} {\bibfnamefont {K.}~\bibnamefont {Hartel}}, \bibinfo {author} {\bibfnamefont {P.}~\bibnamefont {Kienle}}, \bibinfo {author} {\bibfnamefont {H.~J.}\ \bibnamefont {K{\"o}rner}}, \bibinfo {author} {\bibfnamefont {R.~E.}\ \bibnamefont {Segel}}, \bibinfo {author} {\bibfnamefont {P.}~\bibnamefont {Sperr}}, \ and\ \bibinfo {author} {\bibfnamefont {W.}~\bibnamefont {Wagner}},\ }\href@noop {} {\bibfield  {journal} {\bibinfo  {journal} {Zeitschrift f{\"u}r Physik A Atoms and Nuclei}\ }\textbf {\bibinfo {volume} {293}},\ \bibinfo {pages} {119} (\bibinfo {year} {1979})}\BibitemShut {NoStop}%
\bibitem [{\citenamefont {Sch{\"u}ll}\ \emph {et~al.}(1981)\citenamefont {Sch{\"u}ll}, \citenamefont {Shen}, \citenamefont {Freiesleben}, \citenamefont {Bock}, \citenamefont {Busch}, \citenamefont {Bangert}, \citenamefont {Pfeffer},\ and\ \citenamefont {P{\"u}hlhofer}}]{schull1981influence}%
  \BibitemOpen
  \bibfield  {author} {\bibinfo {author} {\bibfnamefont {D.}~\bibnamefont {Sch{\"u}ll}}, \bibinfo {author} {\bibfnamefont {W.}~\bibnamefont {Shen}}, \bibinfo {author} {\bibfnamefont {H.}~\bibnamefont {Freiesleben}}, \bibinfo {author} {\bibfnamefont {R.}~\bibnamefont {Bock}}, \bibinfo {author} {\bibfnamefont {F.}~\bibnamefont {Busch}}, \bibinfo {author} {\bibfnamefont {D.}~\bibnamefont {Bangert}}, \bibinfo {author} {\bibfnamefont {W.}~\bibnamefont {Pfeffer}}, \ and\ \bibinfo {author} {\bibfnamefont {F.}~\bibnamefont {P{\"u}hlhofer}},\ }\href@noop {} {\bibfield  {journal} {\bibinfo  {journal} {Physics Letters B}\ }\textbf {\bibinfo {volume} {102}},\ \bibinfo {pages} {116} (\bibinfo {year} {1981})}\BibitemShut {NoStop}%
\bibitem [{\citenamefont {Wang}\ and\ \citenamefont {Guo}(2016)}]{wang2016new}%
  \BibitemOpen
  \bibfield  {author} {\bibinfo {author} {\bibfnamefont {N.}~\bibnamefont {Wang}}\ and\ \bibinfo {author} {\bibfnamefont {L.}~\bibnamefont {Guo}},\ }\href@noop {} {\bibfield  {journal} {\bibinfo  {journal} {Physics Letters B}\ }\textbf {\bibinfo {volume} {760}},\ \bibinfo {pages} {236} (\bibinfo {year} {2016})}\BibitemShut {NoStop}%
\bibitem [{\citenamefont {Li}\ \emph {et~al.}(2019)\citenamefont {Li}, \citenamefont {Sokhna}, \citenamefont {Xu}, \citenamefont {Li}, \citenamefont {Zhang}, \citenamefont {Li}, \citenamefont {Ge},\ and\ \citenamefont {Zhang}}]{li2019isospin}%
  \BibitemOpen
  \bibfield  {author} {\bibinfo {author} {\bibfnamefont {C.}~\bibnamefont {Li}}, \bibinfo {author} {\bibfnamefont {C.~A.~T.}\ \bibnamefont {Sokhna}}, \bibinfo {author} {\bibfnamefont {X.}~\bibnamefont {Xu}}, \bibinfo {author} {\bibfnamefont {J.}~\bibnamefont {Li}}, \bibinfo {author} {\bibfnamefont {G.}~\bibnamefont {Zhang}}, \bibinfo {author} {\bibfnamefont {B.}~\bibnamefont {Li}}, \bibinfo {author} {\bibfnamefont {Z.}~\bibnamefont {Ge}}, \ and\ \bibinfo {author} {\bibfnamefont {F.-S.}\ \bibnamefont {Zhang}},\ }\href@noop {} {\bibfield  {journal} {\bibinfo  {journal} {Physical Review C}\ }\textbf {\bibinfo {volume} {99}},\ \bibinfo {pages} {034619} (\bibinfo {year} {2019})}\BibitemShut {NoStop}%
\bibitem [{\citenamefont {Jiang}\ \emph {et~al.}(2010)\citenamefont {Jiang}, \citenamefont {Wang}, \citenamefont {Li},\ and\ \citenamefont {Scheid}}]{jiang2010dynamical}%
  \BibitemOpen
  \bibfield  {author} {\bibinfo {author} {\bibfnamefont {Y.}~\bibnamefont {Jiang}}, \bibinfo {author} {\bibfnamefont {N.}~\bibnamefont {Wang}}, \bibinfo {author} {\bibfnamefont {Z.}~\bibnamefont {Li}}, \ and\ \bibinfo {author} {\bibfnamefont {W.}~\bibnamefont {Scheid}},\ }\href@noop {} {\bibfield  {journal} {\bibinfo  {journal} {Physical Review C}\ }\textbf {\bibinfo {volume} {81}},\ \bibinfo {pages} {044602} (\bibinfo {year} {2010})}\BibitemShut {NoStop}%
\end{thebibliography}%

\end{document}